\documentstyle[aps,prl,amstex,twocolumn,epsfig]{revtex}

\begin{document}

\draft
\title{Molecular configurations in a droplet detachment process of a complex liquid}

\author{R.Sattler$^{1}$,  A.~Kityk$^{1,2}$ and C.~Wagner$^{1*}$}
\address{$^1$ Experimentalphysik, Universit\"at des Saarlandes, Postfach 151150, 66041
Saarbr\"ucken, Germany  \\ $^2$ Institute for Computer Science,
Technical University of Czestochowa, Armii Krajowej 17, PL 42-200
Czestochowa, Poland\\ $^*$ c.wagner at mx.uni-saarland.de}
\maketitle

\begin{abstract}
We studied the microscopic polymer conformations in the droplet
detachment process of an elastic semi-dilute polyelectrolytic
Xanthan solution by measuring the instantaneous birefringence. As in
earlier studies, we observe the suppression of the finite time
singularity of the pinch-off process and the occurrence of an
elastic filament. Our microscopic measurements reveal that the
relatively stiff Xanthan molecules are already significantly
pre-stretched to about 90 $\%$ of their final extension at the
moment the filament appears. At later stages of the detachment
process, we find evidence of a concentration enhancement due to the
elongational flow.
\end{abstract}



\vskip2pc

\section{\label{sec:level1}Introduction}

The addition of a tiny amount of polymer to a simple liquid alters
the dynamics of a droplet detachment process dramatically. Instead
of the finite time singularity of the minimum neck diameter
\cite{Eggers97}, a cylindrical filament is formed and the shrinking
dynamics can be slowed down by several
decades\cite{Goldin69,Bazilevskii81,Amarouchene01,Smolka06}. Despite
the technological relevance of the droplet forming process of
complex liquids that reaches from plotting of DNA-microarrays to
food processing, only little is known on the underlying physical
mechanisms. Besides the complexity of the problem and the lack of a
universally applicable constitutive equation for complex liquids,
there exist no microscopic data on the molecular conformation in
such a flow. It is the goal of this study to close this gap.

Earlier experimental studies on the droplet detachment process of
complex liquids were limited to the analysis of macroscopic
quantities like the determination of the shape of the thinning
filament. These can be compared with theoretical analysis on
capillary break-up based on phenomenological polymer models like
e.g. the Oldryd-B or models that follow from kinetic theory like
FENE-P \cite{Li03}. But for a true comparison with microscopic
models, the dynamics of the polymers on the molecular level must
be measured. Birefringence is a typical tool for such
investigations \cite{Fuller}, and we use a high speed set-up
capable of capturing dynamics on the time scale of milliseconds to
study both the microscopic conformation and the macroscopic flow
response in the elastic thread simultaneously.

\section{Sample characterization}

The polymeric system of our study, Xanthan (Sigma-Aldrich), was
chosen because of its high optical anisotropy.  The molecular weight
of t our polymer $m$ is vaguely specified by the distributor with
``$m=2\times10^{6}$ \textit{amu} or more``. Xanthan shows a
pronounced shear thinning and biological digestibility and, for both
reasons, it is widely used in the food, pharmaceutical, oil, and
cosmetic industries. The ordered molecule exists in solution as a
semi rigid helix with a persistence length of 120 nm. It undergoes a
conformational transition to a disordered flexible coil only above a
temperature of $50^\circ C$ \cite{Koenderink04,holzwarth77}.

\begin{figure}
\epsfig{file=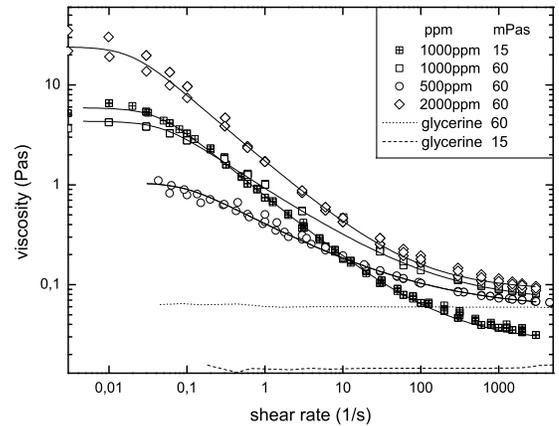, angle=0, width=1\linewidth}
\caption{Stationary viscosities from shear rate and stress
controlled measurements. Not all experiments and data points are
shown. The lines are approximations by the Carreau model.}
\label{fig1}
\end{figure}

Our sample solutions were prepared with weight concentrations
$c_p$ in the semi-dilute regime from $250 ppm < c_p < 2000 ppm$,
mostly dissolved in $80/20$ weight$\%$ glycerol-water as a solvent
with a viscosity of $\eta_{solv} = 60 mPas$. We also used
different glycerol-water weight ratios resulting in solvent
viscosities down to $\eta_{solv} = 2 mPas$ only to allow for a
better determination of the molecular weight of our sample by
fitting our extensional rheological data. The polymer
concentrations where chosen to obtain a sufficiently high
birefringence but still to be significantly below concentrations
at which spontaneous lyotropic ordering is expected in equilibrium
($c_p>5000ppm$) \cite{Koenderink04}. Standard rheological
measurements, using a cone plate geometry on a standard Rheometer
(Haake Mars, Thermo Electron, Germany) were performed to
characterize the shear thinning behavior (fig. \ref{fig1}) of our
samples. Stationary experiments controlling either shear rate or
tension were repeated several times for several samples in a cone
plate geometry with an angle of $0.5^\circ$ and a diameter of $60
mm$. Depending on the experimental deviations an average of
between 6 and 10 measurements is shown. The solutions show a
pronounced shear thinning, as to be expected for relatively stiff
molecules approaching a stationary value slightly enhanced
compared to the pure Newtonian solvent according to the polymer
concentration. The data have been approximated using the Carreau
model \cite{tanner85} and the results are in reasonable agreement
with ref. \cite{lopez02} if one takes our higher solvent
viscosities into account. The characteristic relaxation times of
the fits range from about $10 s$ for $500 ppm$ and $20 s$ for
$1000 ppm$ to about $50 s$ for the $2000 ppm$ solution. These time
scales give a measure for the rotational diffusion time of our
molecules and must be compared with the elongational rates in our
droplet experiments.

For the later discussion of our elongational viscosity
measurements we would like especially to point out that the 1000
ppm samples with different solvent viscosities show similar zero
shear rate viscosities, and the differences only become obvious at
high shear rates.

\section{Experimental setup}

\begin{figure}
\epsfig{file=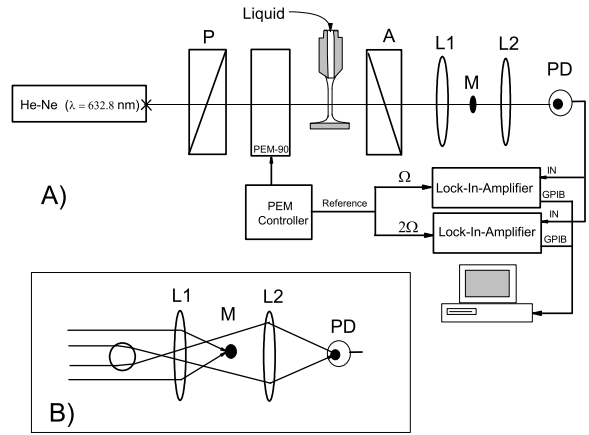, angle=0, width=1\linewidth}
\caption{The experimental setup. See text for further
explanation.} \label{fig2}
\end{figure}

\begin{figure}
\epsfig{file=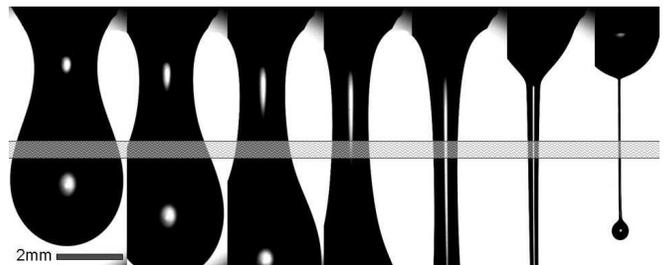, angle=0, width=1\linewidth}
\caption{Photographs of the droplet detachment process of a 2000
ppm Xanthane solution ($\eta_{solv}$=60mPas) taken every 0.1s
indicating, schematically, the laser beam positioned close to the
neck and in a region as cylindrical as possible throughout the
measurement.} \label{fig3}
\end{figure}

The experimental setup is sketched in fig. \ref{fig2}A. The polymer
solutions are quasistatically driven through a nozzle of $d = 2 mm$
diameter by a syringe pump. At a distance of $D = 7 cm$ below the
nozzle, a plate is mounted in order to omit gravitational effects,
and to stabilize the filament against air currents. All measurements
have been performed at room temperature ($25^\circ C$). The
retardation $\delta$ is measured using an optical train consisting
of an 18mW HeNe Laser (JDS Uniphase), a Polarizer P, a Photo Elastic
Modulator PEM (Hinds Instruments), the sample liquid, an Analyzer A,
a bright field lens system L1 and L2, and a photo diode PD connected
to two Lock-Ins (Stanford Research). The polarizer is fixed at an
angle of $45^\circ$ relative to the photo elastic modulator which
allows for a Lock-In technique for the detection of the retardation
signal. The bright field lens system (fig. \ref{fig2}B) in front of
the detector prevents any light that has not crossed the filament
(fig. \ref{fig3})) from reaching the detector by blocking the
parallel components with a mask M in the focal point of lens L1. The
lens L2 collects the light that is diffused by the cylindrical
filament into the detector \cite{Talbot79}. To calculate the
birefringence $\Delta n = \delta h(t)/(2 \pi \lambda) $ (with
$\lambda = 633nm$ as the HeNe laser wavelength) from the retardation
signal the instantaneous thickness $h(t)$ of the filament must be
known. The thinning process of the filament diameter $h(t)$ is
measured at 500 frames per second with a high speed CCD-camera
(encore mac PCI 1000S) that is placed perpendicular to the optical
train. The camera is equipped with a 2 times magnification objective
and the image capturing is synchronized to the data collection of
the two Lock-In´s. From fig. \ref{fig3} it becomes obvious that, for
geometric reasons, it is impossible to measure birefringence before
the droplet has completely passed the light beam and the temporal
exponentially thinning filament is present. This is the moment when
our measurements start.

\section{Experimental results}
\subsection{Macroscopic measurements}

Different regimes can be distinguished analysing the temporal
behavior of the neck diameter $h(t)$ in the droplet detachment
process of a polymer solution. First, it follows an exponential
law that is well described by the Rayleigh instability of
Newtonian liquids \cite{Rothert01,Wagner05}. Then it might pass
into the self similar regimes that describe the break-up of
Newtonian liquids \cite{Eggers93} until the polymers intervene the
flow abruptly and a filament is formed which, again, thins very
slowly and exponentially with time \cite{Yarin84}.

\begin{figure}
\epsfig{file=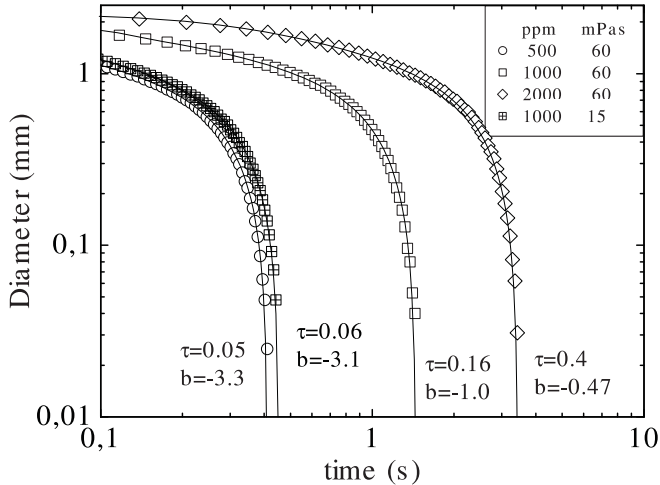, angle=0, width=1\linewidth}
\caption{Filament diameter h(t) for different polymer
concentrations and solvent viscosities. The parameters refer to
eq. \ref{fit}. Not all data points are shown.} \label{fig4}
\end{figure}

\begin{figure}
\epsfig{file=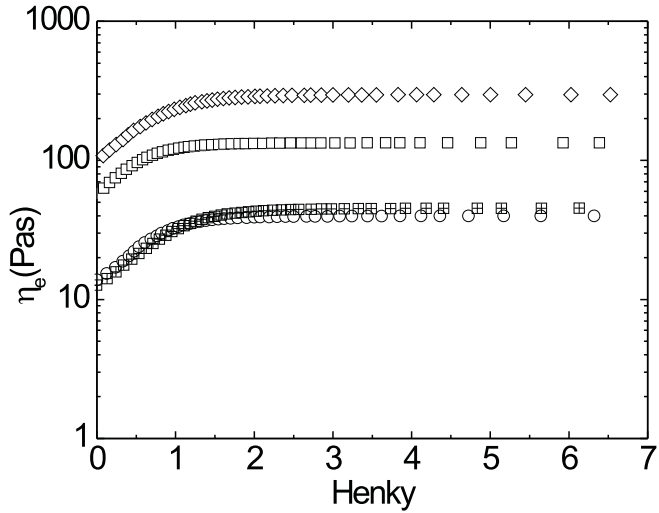, angle=0, width=1\linewidth} \vspace{0.5cm}
\caption{The apparent elongational viscosity for the same set of
data as in fig. \ref{fig4} } \label{fig5}
\end{figure}

Figure \ref{fig4} shows the filament thickness $h(t)$ extracted from
the video images that start at the moment when the filament is
formed. More viscous solvents or higher polymer concentrations lead
to a slower dynamic that differs by more than an order of magnitude.
However, at first all runs show the same exponential thinning
behavior, followed by a linear regime, and the data can be
approximated by the expression
\begin{equation}
h(t)=exp(-t/\tau)-bt+y_o.
\label{fit}
\end{equation}
The formation of the cylindrical filament occurs due to the high
resistance of the polymers to elongational flow and this resistance
is macroscopically quantified by the elongational viscosity
$\eta_e$. The same form of equation \ref{fit} has been used to fit
data from extensional rheology measurements (both capillary break up
and filament stretching devices\cite{Anna00}). Its physical meaning
is that first the polymer molecules are uncoiled by the flow and the
dynamic is governed purely by elastic effects. Once the polymers are
stretched, a steady state is reached which leads to the linear
dynamic. From  $h(t)$ one can calculate the corresponding
elongational rates $\dot{\epsilon}=\partial_th(t)/h(t)$. These have
to be compared with characteristic relaxation times. The microscopic
polymer rotational diffusion has been estimated by the time
constants from the standard rheological measurements to be at least
$\lambda
> 10 s$ and the appropriate ratio is given by the Weissenberg
number which is always found to be very large $Wi=\lambda
\dot{\epsilon} > 100$ in our experiments. This means that the
polymer molecules in the droplet detachment process do not
experience a force field that is comparable to the Brownian forces
and time scales, but are affected by a violent flow and stretched
and oriented in a deterministic manner. Nevertheless, the
elongational rates and the Weissenberg numbers are practically
constant in the regime of exponential thinning. They diverge,
though, when the linear thinning behavior sets in.

The $h(t)$ data can also be used directly to calculate the
apparent elongational viscosity  by equating the capillary
pressure with the elastic stresses
\cite{Bazilevskii81,Amarouchene01,Anna00}:

\begin{equation}
\eta_e(t) =  \frac{2\sigma}{\dot\epsilon h(t)} \label{elovisco}
\end{equation}

This approach is a simplification because it neglects
gravitational or nonlinear elastic effects, but it is commonly
used to extract the elongational viscosity from capillary break up
experiments. A common quantity representing the stress- and
stretching history of the fluid, and thereby the polymers, is the
Henky strain given by $\zeta = \int\dot{\epsilon} dt$. The
corresponding elongational viscosity data in fig. \ref{fig5} first
show an exponential increase of the apparent elongational
viscosity followed by a steady state value for strains $\zeta \geq
1.5$. The observation of the plateau value at strains that much
smaller compared to earlier work
\cite{Amarouchene01,Anna00,Wagner03} is a consequence of the
relative stiffness of the Xanthan molecule and the pre-stretching
before the occurrence of the filament. This is also indicated by
the elevated elongational viscosity at $\zeta = 0$. If the
filament would start to be formed with the polymers at rest one
would expect the elongational viscosity to grow from the Newtonian
value of the so called Trouton ratio $\eta_e/\eta_s=3$. When
$\eta_e$ reaches the plateau value the thinning dynamic becomes
linear and the elongational rate $\dot{\epsilon}$ diverges,
indicating that $\eta_e$ is independent of the elongational rate
as predicted by rigid rod models as well as elastic models for
$Wi>>1$ \cite{Bird96}. The measured elongational viscosities are
in reasonable agreement with data from studies with opposing
nozzles \cite{Fuller87}, fiber spinning devices \cite{Khagram83},
or values obtained by analyzing contraction flow \cite{Mongruel03}

\begin{figure}
\epsfig{file=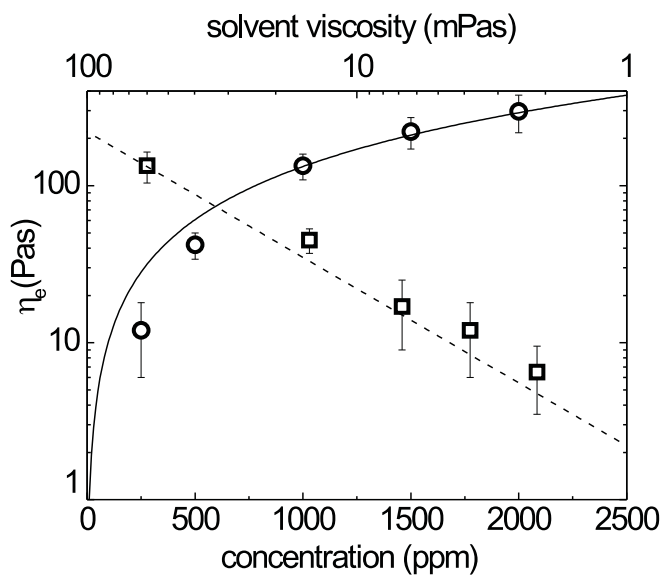, angle=0, width=1\linewidth}
\caption{The elongational viscosity for different solvent
viscosities (squares) and polymer concentrations (circles). The
continuous and the dashed lines show the theoretical values
according to Batchelors formula for semidiluted rigid rods. }
\label{fig6}
\end{figure}

For our birefringence study we needed an estimate for the molecular
weight of our sample and we used Batchelors formula \cite{Batchelor}
for the elongational viscosity of semidiluted rigid rods to fit our
data. Fluorescence microscopy studies on DNA show that at steady
state for Henky strains $\zeta \geq 1.5$ the molecules are extended
to a maximum hydrodynamic length $\ell$ that is more than $80\%$ of
their contour length \cite{Smith98}, and the relatively stiff
Xanthan molecules are mostly well approximated as rigid rods. In
order to allow for a robust fitting procedure for the determination
of the two free parameters molecular weight $m$ and polymer length
$\ell$ we varied both the polymer concentration and the solvent
viscosity. As expected we found the elongational viscosity to always
be proportional to the solvent viscosity (see fig. \ref{fig5}). With
our two independent sets of data, we performed the fit with
Batchelors formula for semidiluted solutions of rigid rods (fig.
\ref{fig6})

\begin{equation}
\eta_e =3\eta_{solv}+\frac{\frac{2}{3}\eta_{solv}\Phi(\ell/d)^2}
{ln(2\ell/d)-ln(1+2\sqrt{\Phi/\pi}\ell/d)-1.5}\label{Batchelor}
\end{equation}

$\Phi$ is the particle volume fraction and, for the width of the
polymer, we took the literature value of $d = 2nm$. The fit yields
$\ell=4.5 \mu m$ and $m = 2.5$ M\textit{amu}. This ratio of
molecular weight and length is in very good agreement with e.g.
the results from a contraction flow study by \cite{Mongruel03},
but differs by a factor of two from values obtained by light
scattering \cite{Fuller87}. The number of Kuhn steps of our sample
is about $40$, a value that is 10 times smaller than that of
$\lambda$-DNA which has been used in a previous study where no
steady state of the elongational viscosity had been observed
\cite{Wagner03}.

\subsection{Birefringence measurements}

We can now turn to the microscopic study of the conformation of
the macromolecules by means of the birefringence measurements. An
elaborate introduction in optical flow rheometry is given in
\cite{Fuller}. In our unidirectional stretching geometry with the
flow along the direction of gravity, we can assume transverse
isotropy for polarizability and segmental order of the single
units of the macromolecules. Therefore, the birefringence signal
$\Delta N$ is proportional to the segmental order of the polymers
and to their concentration
\begin{equation}
\Delta N = \frac{2 \pi}{n m}\Delta\alpha_0 c_p N_a
\frac{1}{2}<3\cos^2\theta-1>
\end{equation}
where n is the average refractive index, $N_a$ the Avogadro number
and $\theta$ the orientation angle of the polymer segments to the
direction of extension. By knowledge of the maximum birefringence
signal $\Delta N_{max}$ at maximum extension and alignment of the
polymers, the mean fractional extension of our polymers is then
$x/\ell=\sqrt{ (2\Delta N/\Delta N_{max} + 1)/ 3}$ because, on a
microscopic scale, a variety of different configurations will
certainly be present \cite{Smith98}.

The negligence of different configuration types like, e.g.,
dumbbell or hairpin, leads to a certain overestimate of the
fractional extension and our calculation looses its accuracy for
coiled states. The correctness of the formula improves, though,
increasing degree of orientational order and is a good
approximation for large fractional extensions of the stretched
polymers.

\begin{figure}
\epsfig{file=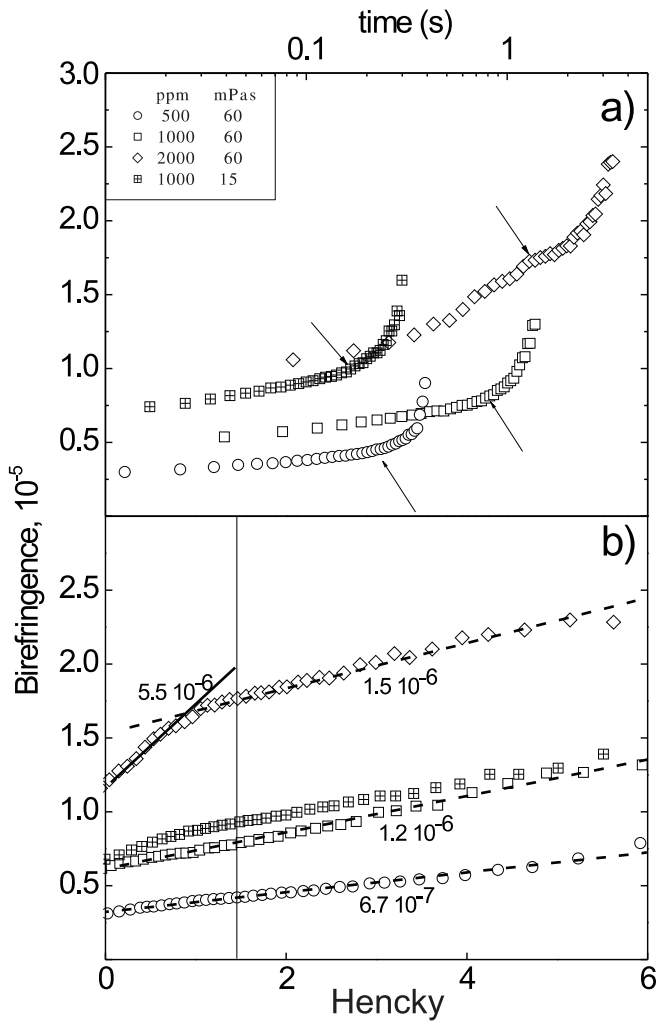, angle=0, width=1\linewidth}
\caption{a) The instantaneous birefringence versus the elapsed
time. The arrows indicate a Hencky strain of $\zeta=1.5$  b) the
instantaneous birefringence versus the Hencky strain. The vertical
line indicates the $\zeta=1.5$ position, the dashed lines are
linear fits with slopes indicated by the numbers. Symbols are as
in fig. \ref{fig1}. } \label{fig7}
\end{figure}

To test the reliability of our method we performed several runs at
different polymer concentrations, and the birefringence signal was
always roughly proportional to the concentration (fig.
\ref{fig7}b). The robustness and selectivity of our method becomes
clear if one compares the data of the solution with $500 ppm$ of
polymer and $\eta_{solvent} = 60 mPas$ with the solution with
$1000 ppm$ of polymer and $\eta_{solvent} = 15 mPas$. Both data
sets practically lead to the same flow (fig. \ref{fig4}) but,
exceeding the capabilities of simple extensional rheometry, the
birefringence signal is sensitive to the differences in
concentration. This shows how important the microscopic
measurements are if one wants to qualify theoretical predictions
that are based on kinetic models.

The measurements of the birefringence in our setup are restricted
to the time $t_c$ after the filament is formed because optical
abberations make it impossible to measure it at earlier stages of
the pinching process. Even if the origin of the Hencky strain is
not influenced much by this limitation, as most of the strain is
accumulated at the final stages of the experiment, our data reveal
that the polymers have a significant (flow and strain) history
when our measurements start. Assuming that the polymers are
completely stretched when elastic effects are no longer observable
for $\zeta \geq 1.5$, we can calculate the maximum birefringence
per polymer concentration $\Delta N_{max}/c_p = 8\times 10^{-2}
g^{-1}cm^3$. This is comparable to the value for fully elongated
$\lambda$ DNA of $\Delta N_{max}/c_p = 5.1\times10^{-2}
g^{-1}cm^3$ \cite{Maret83}, but a decade larger than values
obtained for Xanthan in shear flow \cite{Yevlampieva99} and in
transient electric birefringence measurements \cite{Morris82}.

The corresponding maximum birefringence of the $c_p \leq 1000 ppm$
solutions is only $~30\%$ larger than at $\zeta = 0$. This would
correspond to a fractional polymer extension $x/\ell = 0.9$ at
$\zeta = 0$, indicating that stiff polymers like Xanthan are almost
completely aligned and stretched when the filament occurs. The
uncoiling process in the range $0<\zeta< 1.5$ correlates with an
elastic, exponential filament thinning dynamic and a monotonic
increase of the birefringence signal.

However, instead of a plateau value at the final stage of the
experiment, we observe a divergence in the birefringence signal
(fig. \ref{fig7}a). The divergence is very reproducible and takes
place at stages of the filament thinning process when the filament
diameter is still large enough to render the analysis unambiguous.
The divergence of the birefringence signal comes along with a
divergence in the elongational rate $\dot{\epsilon}$ that follows
from the linear shrinking of the filament. Surprisingly, we find
that, the birefringence signal continues to increase with constant
slope from $\zeta = 1.5$ to $\zeta = 6$ if plotted against the
strain. Only for the $c_p \geq 1500 ppm$ solutions did we observe
a steeper slope in the birefringence signal for $\zeta < 1.5$,
presumably because of the stronger entanglement of the polymers.

While the polymers should be completely uncoiled in the linear
thinning regime, additional physical mechanisms, that we will
discuss in the following, have to be considered. We do see
\textit{three} different possible scenarios: \textit{first} an
effect of the polydispersity of the molecular weight distribution of
our sample. But the observed large Wi should lead to a complete
stretching of smaller molecules at even smaller strains.
\textit{Second} an overstretching of the molecule by the strong flow
that would affect the configurations of the molecular bonds.
\textit{Third} a concentration enhancement by drainage or
evaporation of the solvent. This picture is supported by the
observation that, eventually, the filament might not break at the
final stages of the thinning process but leave a very thin polymer
($<10\mu m$) fiber between the nozzle and the ground plate. Then a
configuration called beads on a string occurs via an instability of
the surface of the cylindrical filament \cite{mckinley05}.

\section{Conclusion}

In conclusion, we have presented the first measurements on the
molecular configurations of semi-rigid Xanthan molecules in a
droplet detachment process of a complex liquid. We find that stiff
molecules like Xanthan are highly oriented and stretched by the
flow, even \textit{before} the abrupt transition from the Newtonian
self similar shrinking law to the exponential behavior of the
elastic filament. Furthermore, we do observe a further increase of
the birefringence signal after the saturation of elastic effects and
the stretching of the polymers. We discussed different possible
reasons for this phenomenon and find that a concentration
enhancement is most likely to be the pertinent effect, but we cannot
exclude the possibility that changes in the intramolecular bond
configurations caused by the violent flow affect the birefringence
signal too.

\end{document}